\documentclass[a4paper,11pt]{article}
\usepackage[margin=1in]{geometry}
\usepackage{epsfig}
\usepackage{cite}
\usepackage{graphicx}
\usepackage{epstopdf}
\usepackage{color}
\UseRawInputEncoding
\usepackage[nottoc]{tocbibind}
\usepackage[english]{babel}
\usepackage{tabularx}
\usepackage{xparse}
\usepackage{braket}
\usepackage{amsmath}
\usepackage{hyperref}
\usepackage{grfext}
\usepackage{amssymb}
\usepackage{caption}
\usepackage{mathtools}
\usepackage{subcaption}
\usepackage{algorithm2e}
\usepackage{float}
\usepackage{subcaption}
\usepackage{multirow}
\restylefloat{table}
\usepackage{booktabs}
\usepackage[dvipsnames]{xcolor}
\usepackage{tikz}
\usepackage{pdflscape}
\usepackage{lscape}
\date{\today}
 \normalsize

\newcommand{\bmat}{\left(\begin{array}}
\newcommand{\emat}{\end{array}\right)}
\newcommand{\be}{\begin{equation}}
\newcommand{\ee}{\end{equation}}
\newcommand{\bea}{\begin{eqnarray}}
\newcommand{\eea}{\end{eqnarray}}

\def\gtwid{\mathrel{\raise.3ex\hbox{$>$\kern-.75em\lower1ex\hbox{$\sim$}}}}
\def\ltwid{\mathrel{\raise.3ex\hbox{$<$\kern-.75em\lower1ex\hbox{$\sim$}}}}
\def\gev{{\rm \, Ge\kern-0.125em V}}
\def\tev{{\rm \, Te\kern-0.125em V}}

\def    \be            {\begin{equation}}
\def    \ee            {\end{equation}}
\def    \bea           {\begin{eqnarray}}
\def    \eea           {\end{eqnarray}}

\def\a{\alpha}
\def\b{\beta}

\def\d{\delta}
\def\n{\nu}

\def\m{\mu}

\def\d{\delta}

\def\gmu{g_{\mu\nu}}
\def\half{\frac{1}{2}}

\def\m{\mu}

\def\d{\delta}

 \normalsize

\begin{document}

\vspace{.3cm}

\title{\Large \bf Inflation by Variation of the Strong Coupling Constant: update for Planck 2018}
\author
{  \it \bf M. AlHallak$^{1}$\thanks{phy.halak@hotmail.com},  A. AlRakik$^{1}$\thanks{alrakikamer@gmail.com}, S. Bitar$^{2}$\thanks{shafee.bitar@hiast.edu.sy}, N. Chamoun$^{2}$\thanks{nidal.chamoun@hiast.edu.sy} and M. S. Eldaher$^{3}$\thanks{eldaherm@gmail.com}
 \\\hspace{-3.cm}
\footnotesize$^1$  Physics Dept., Damascus University, Damascus, Syria  \\\hspace{-3.cm}
\footnotesize$^2$ HIAST, P.O. Box 31983, Damascus, Syria,
\\\hspace{-3.cm}
\footnotesize$^3$ Higher Institute for Laser Research \& Applications, Damascus University, Damascus, Syria.
}

\date{}
\maketitle

\abstract{We apply the ``systematic" $1^{st}$ order cosmological perturbation theory method to re-derive the formulation of an inflationary model generated by variation of constants, then to study the case where it is non-minimally coupled to gravity within both the ``Metric" and ``Palatini" formulations. Accommodating Planck 2018 data with a length scale $\ell$ larger than Planck Length $L_{pl}$ requires amending the model. First, we assume $f(R)$ gravity where we show that an $R^2$-term within Palatini formulation is able to make the model viable. All along the discussions, we elucidate the origin of the difference between the ``Metric'' and "Palatini" formalisms, and also highlight the terms dropped when applying the shortcut ``potential formulae method", unlike the ``systematic' method", for the observable parameters. Second, another variant of the model, represented by a two-exponentials potential, fits also the data with $\ell> L_{pl}$.}

{\bf Keywords:} Variation of Constants, Inflation

{\bf PACS:} 98.80Cq, 98.80-k,

\section{Introduction}

 Inflation  \cite{Guth} is the commonly accepted theory for solving many of the Big Bang scenario problems, mainly the horizon and flatness problems. Most of inflationary models introduce a scalar field, the inflaton, responsible for the inflation, whose nature is not well understood \cite{ecyclopedia-inflation}. Many models differ in the details, but most attribute a matter content to the inflaton. Varying speed of light (VSL), \cite{magueijo} was an alternative for solving the Big Bang scenario problems. VSL is nonetheless a part of ``variation of constants'' ideas \cite{vucetich-landau}. Experimental data preclude any temporal variation of the electric charge \cite{bekenstein}, the strong coupling \cite{chamoun-PLB} and the Higgs vacuum expectation value (vev) \cite{chamoun-JPhG},  going back in time till nucleosynthesis. However, no data exist to preclude variation of constants in the inflationary era. In \cite{chamounIJMPD}, a link was suggested between variation of constants and inflation, and in \cite{chamounJCAP} a detailed study of an inflationary model generated by the temporal variation of the strong coupling constant was presented and contrasted to Planck 2015 data. In some scenarios involving multiple fields, the model was viable with data.

For slow-roll inflation, one can use the $1^{st}$- order cosmological perturbation theory, henceforth called ``systematic method", in order to evaluate the spectral index $(n_s)$ and the tensor-to-scalar ratio $(r)$ \cite{Mukhanov}. There is a shortcut, which we call the ``potential method", where the inflation parameters ($n_s,r$) are given in terms of the potential and its derivatives, provided this potential is expressed in terms of ``canonical" fields minimally coupled to gravity and with standard kinetic energy terms \cite{Liddle}.

We do not advocate one particular method, but rather use one to crosscheck the other. The ``potential method" is simple but usually requires a non-trivial transformation in order to put the Lagrangian in ``canonical" form, whereas the ``systematic method" is more general but less straightforward. Thus, if the Lagrangian involves a complicated non-decomposable function $P(\phi, \partial^\mu \phi \partial_\mu \phi)$ of the matter field and its kinetic energy term, then it is not possible to transform it into another frame where the Lagrangian becomes canonical, and here one can only apply the ``systematic method" \cite{Main-1103.1172}.

The simple model of varying $\a_S$ was examined in \cite{chamounJCAP} using the ``potential method", whereas in this letter we start by re-deriving it using the ``systematic method", and show its non-viability with data.

Then, we extend the model by assuming a variant with a non-minimal coupling to gravity (non-MCtG) \cite{Demir-0803.2664}. Here, one can use two physically distinct formulations: the ``Metric" and the ``Palatini" formalisms. In the latter, the Christoffel connection is considered independent of the metric, whereas in the former, one uses the Levi-Civita connection defined in terms of the metric. Thus, in the ``systematic method", the starting point Einstein equations, obtained by perturbing the metric, are different in the two formulations, since $R_{\m\n}$ remains fixed in the ``Palatini'' formulation, in contrast to the ``Metric" one. However, the equation obtained by varying the connection in the Palatini approach leads to it being equal to the Levi-Civita one, since the Hilbert-Einstein action contains only Ricci scalar $R$.

In both ``Metric" and ``Palatini" formulations, one usually does a conformal transformation on the metric in order to move from Jordan frame to Einstein frame. In the latter frame, one can use the ``potential formulae" noting however that the transformation from the old field to the new `canonical' field is different in the two formulations. In ``Palatini", one can invert the transformation and express the old field in terms of the new one, thus reaching a closed form of the `canonical' potential. On the other hand, one resorts usually to numerics in order to determine the Einstein frame potential in the ``Metric" formulation.

We checked that both the ``systematic" and ``potential" methods give identical results for the ``Metric" and ``Palatini" formulations, and in both cases, the model, despite showing some improvement compared to the minimal coupling to gravity case, is still unable to fit the data.

In order to make the model viable, we add a new ingredient by assuming $f(R)$ gravity. More precisely, we assume a Hilbert-Einstein Lagrangian density of the form ($R+\a R^2$) \cite{faraoni-0805.1726}. Note here that in ``Palatini" formulation, the connection $\Gamma$ is not the Levi-Civita one, and so the Palatini Ricci scalar is different from the ``Matric" Ricci scalar. For the ``systematic method", one can a priori find the corresponding Einstein equations in the original Jordan frame by perturbing the metric and see the contributions of the additional term ($\a R^2$) keeping only those of $1^{st}$ order in metric perturbations. One anticipates, though, obtaining complicated formulaes originating from $(R+\d R)^2$ in both the ``Metric" and ``Palatini" formulations. That is why we follow the usual way of treatment of $f(R)$ by trading off $R$ with an auxiliary field $\varphi$ in addition to the original `matter' field $h$. This scalar field $\varphi$ is dynamical in the ``Metric" formulation and represents a genuine degree of freedom, which would show itself upon moving to the `canonical' Einstein frame by the presence of a kinetic term $(\partial \varphi \partial \varphi)$, so its Euler-Lagrange equation cannot be solved algebraically.

For simplicity, thus, we adopt the ``Palatini" formulation of $f(R)$, where the auxiliary field $\varphi$ is not dynamic and its Euler-Lagrange equation can be solved algebraically to express $\varphi$ in terms of $(h, \partial h)$. Upon moving to the `canonical' Einstein frame by transforming $h$ into $\chi$ with canonical kinetic energy term, we see that the $(\a R^2)$ term would show itself in two effects: the first is to have an effective potential $U(\chi)$ including an $\a$-term, whereas the second is the presence of a higher kinetic term $\a (\partial \chi \partial \chi)^2$.

The ``systematic method" can be applied at this stage by perturbing the metric $(g_{\m \n} \rightarrow g_{\m \n} + \d g_{\m \n})$. The term $\a (\partial \chi \partial \chi)^2$ would contribute in quantities of $1^{st}$ order in the metric perturbation. However, for small $\a$, the leading term $\a(\partial_\m \chi \partial^\m \chi)^2$ would be subdominant during the slow-roll inflationary era compared to the kinetic term $(\partial_\m \chi \partial^\m \chi)$, whereas the other term involving $(\a \d g^{\m \n} \partial_\m \chi \partial_\n \chi)$ would be again neglected compared to the $1^{st}$-order $\a$-correction in $U(\chi)$. In short, restricting to $1^{st}$-order terms in $\a$ and in the metric perturbation $\d g$ would amount just to drop this higher order kinetic term, and the ``systematic method" would be equivalent to the ``potential method" considering only the effective $U(\chi)$.

Applying the latter method, we calculate $(n_s, r)$ and find that the new degree of freedom of $\a R^2$ does not change $n_s$ much, whereas allows to divide $r$ by a term involving the product of $\a$ and the absolute magnitude of the potential, which can be adjusted to decrease $r$ significantly, enabling the accommodation of the updated data
 of Planck 2018 \cite{planck2018}, separately and combined with other experiments \cite{BKP}.


Finally, we test another variant of the model with no $f(R)$, nor non-MCtG, but where the potential is a combination of two exponentials and find that it is consistent with cosmological data while at the same time keeping its spatial scales larger than $L_{pl}$. Inflation generated by such potentials were studied in \cite{cadoni} in the context of instabilities of de Siter space. However, we follow here \cite{chamounJCAP}, and argue that such potentials can be motivated equally by temporal variation of coupling constants corresponding to two gauge groups.

 The plan of the letter is as follows. In section 2 we present our analysis of the original model stating in detail the ``systematic method". In section 3 we study one variant of the model assuming a non-MCtG and we present the results we obtained in both the ``Metric" and the ``Palatini" formulations, and contrast to Planck 2018 and other experimental data. In section 4, we treat the case of ``Palatini" $f(R)$ applied to our model, and test its ability to accommodate the experimental data. In section 5,
 we study the other variant of the model with two-exponentials potential which seems under perturbative control and can accommodate the data. We end up by a summary and conclusion in section 6.

\section{Analysis of the basic model}
\label{1}

Our starting point is the general four dimensional action:
\bea S = S_{EH}+S_{\phi}\eea
where $S_{\phi}$ is the varying strong coupling constant action given by \cite{chamounJCAP}:
\bea \label{action} S_{\phi}\equiv \int d^4x \sqrt{-g} \mathcal{L}_{\phi}  &=& \int d^4x \sqrt{-g} \lbrace -\frac{1}{2} f(\phi) g^{\mu \nu} \partial_\mu \phi \partial_\nu \phi -V(\phi) \rbrace\eea
 where $f(\phi)=\frac{1}{\ell^2 \phi^2}$, and $V(\phi)=\frac{V_0}{\phi^2}$ with $\phi$ embodying the strong coupling constant variation $g(x)=g_0  \phi(x)$ and $\ell$ is the Bekenstein length scale, and $V_0=\frac{\langle G^2\rangle_T}{4}$ encodes the gluon field strength vev at inflation temperature $T$,  whereas  $S_{EH}$ is the usual Einstein-Hilbert action adopting units where the Planck mass $M_{pl}$ is equal to one:
 \bea S_{EH} &=& \int d^4x \sqrt{-g} \lbrace\frac{R}{2} \rbrace\eea
 with $R$  the Ricci Scalar constructed from the metric $g_{\mu\nu}$. Note that the form of the potential is not put by hand, but rather is dictated by the physical assumption of a varying strong coupling constant, where gauge and Lorentz invariance impose this form originating from the gluon condensate \cite{chamounIJMPD, chamounJCAP}.

 Instead of using directly standard formulae involving the potential function and its derivatives in order to compute the observable spectral parameters, we follow in this section a systematic analysis \cite{Main-1103.1172} applied to our model, which would be applicable in other settings involving extra Lagrangian terms in addition to the kinetic and potential ones.

\paragraph{\bf \underline{Background Equations and the Slow Roll Parameters}} By varying with respect to the flat Friedmann-Robertson-Walker (FRW) metric $g_{\mu\nu}=diag(-1,a^2(t),a^2(t),a^2(t))$, we get  the Einstein equations:
\bea \label{EinsteinEqsBack} G_{\mu \nu}= \mathcal{L}_{\phi} g_{\mu\nu}+f(\phi)\partial_{\mu} \phi \partial_\nu \phi \eea
where $G_{\mu\nu}$ is the Einstein tensor.
Assuming a homogenous and isotropic universe with $\phi=\phi(t)$, we can find the modified Friedmann equations by taking the ($00$) and ($11$) components of Eq. (\ref{EinsteinEqsBack}):
\bea \label{Friedmann-eqs} 3 H^2-\half f(\phi) \dot{\phi}^2-V(\phi)=0\ &,&  2  \dot{H}+3  H^2 +\frac{1}{2} f(\phi) \dot{\phi}^2-V(\phi)=0 \eea
As to the $\phi$ equation of motion, we get:
\bea \label{eqofmotionforphi} f(\phi) \ddot{\phi}+3 H f(\phi) \dot{\phi} +\frac{1}{2} f_{,\phi} \dot{\phi}^2 +  V_{,\phi} =0\eea
where $V(f)_{,\phi}=\frac{dV(f)}{d\phi}$.
 From Eq. (\ref{Friedmann-eqs}), we can derive the slowly varying parameters defined by
 \bea \label{slowrollparametr}  \varepsilon \equiv - \frac{\dot{H}}{H^2}= \frac{\half f(\phi) \dot{\phi}^2}{ H^2} &,& \delta \equiv \frac{d \ln \varepsilon}{dN}= \frac{d \ln \varepsilon}{Hdt}=\frac{\dot{\varepsilon}}{H \varepsilon} =\frac{f_{,\phi} \dot{\phi}}{H f}+\frac{2\ddot{\phi}}{H \dot{\phi}}-\frac{2\dot{H}}{H^2}\eea
 where we assume that during inflation the Hubble parameter changes slowly, so the condition $\varepsilon<<1$ is satisfied during a sufficiently large number of Hubble times (in order to solve the horizon problem and get at least $N \sim 40$-$60$ e-folds), a condition which is expressed as $\delta <<1$ during inflation. Using the slow roll parameters to simplify the equations of motion and to calculate the number of e-folding, we find:
\bea \label{slowrollfriedmann} H^2 \approx \frac{V(\phi)}{3} &,&
  3 H f(\phi) \dot{\phi} \approx - V_{,\phi}, \\
\label{Nequation} N \equiv \int_{a_i}^{a_f} d \ln a =\frac{1}{2  \ell^2} \ln \phi |_{\phi_i}^{\phi_f}  &,& \varepsilon \approx 2 \ell^2, \delta \approx 0 \eea
\paragraph{\bf \underline{Perturbations}} The metric of the perturbed universe is given in the general form as
\bea \label{perturbedMetric}  ds^2=-(1+\lambda)dt^2 + 2 B_i dx^i dt +a^2[(1-2\psi)\delta_{ij}+E_{ij}]dx^i dx^j\eea
where: $B_i\equiv\partial_i \beta-S_i$ with $\partial^i S_i =0$ and $E_{ij}\equiv 2\partial_i \partial_j E +2 \partial_{(i} F_{j)}+h_{ij}$ with $\partial^iF_i=0$ and $h_i^i=0$.
\subparagraph{\bf \underline{Scalar Perturbations}}
At first, we consider only the scalar metric perturbations about the flat FRW background, so the  metric of Eq. (\ref{perturbedMetric}) becomes:
\bea \label{scalarperturMetric} ds^2=-(1+\lambda)dt^2+a^2(1-2\psi)dx_i dx^i +2\partial_i \beta dx^idt+2 a^2 \partial_i \partial_j E dx^i dx^j\eea
It is convenient to use the Arnowitt-Desser-Misner(ADM) form with the line element \cite{ADM}:
\bea \label{ADMmetric} ds^2 &=& -(N^2-\gamma^{ij} N_i N_j)dt^2 +2N_i dx^i dt + \gamma_{ij} dx^i dx^j\eea
 where $\gamma_{ij}$ is the metric on the constant-$\gamma$ hypersurface, $N$ is the lapse and $N_i$ is the shift vector to be expanded respectively as: $N=1+\Phi$ and $N_i=\partial_i \tilde{B}$. The gauge is fixed to the uniform-field gauge leading to gauging away the ``$\partial_i \partial_j E$''-part in $\gamma_{ij}$ which is written as:
\bea \label{uniformGauge}  \gamma_{ij}=a^2(t) \exp{(2\Re)}\delta_{ij},
\eea
where $\Re$ is the comoving curvature perturbation. At the linear level, we have:
\bea \label{linearPertMertic}  ds^2=-(1+2\Phi) dt^2 +2 \partial_i \tilde{B} dt dx^i + a^2(t) (1+2 \Re)dx^i dx_i\eea
so at this level we have $\lambda = 2 \Phi, \psi = - \Re, \beta = \tilde{B}$.

By expanding the action up to the second order in the perturbation variable, one finds:
  \bea \label{expandedAciton} \delta S &=& \int dt d^3x  \left[-3 a^3\dot{\Re}^2+2a \dot{\Re}\partial^2\tilde{B}-2 a H \Phi \partial^2\tilde{B}-2a  \Phi \partial^2\Re \right. \nonumber \\ &&\left.
 +6 a^3H\Phi \dot{\Re}-3a^3 H^2 \Phi^2+\frac{a^3f(\phi)\dot{\phi}^2}{2}\Phi^2 +a  \partial_i\Re \partial_i \Re \right],\eea
that would allow to derive the equations of motion for $\Phi$ and $\tilde{B}$:
\bea \label{equMotionOf}  \Phi= \frac{1}{H} \dot{\Re} &,& \partial^2 \tilde{B}=\frac{a^2}{2H^2}f(\phi)\dot{\phi}^2\dot{\Re}-\frac{1}{H}\partial^2 \Re, \eea
which one can substitute back in Eq.(\ref{expandedAciton}) to get the $\Re$ equation of motion:
 \bea \label{equaOfMotionOfRe}  \frac{d}{dt}(a^3 \varepsilon \dot{\Re})-a \varepsilon \partial^2 \Re=0\eea
 where:
 $\varepsilon=-\frac{\dot{H}}{H^2}$ is the slow-roll parameter. We redefine the field $\Re$ now into Mukhanouv variables in order to have a canonical action, so we put $\nu=Z \Re$ where $Z=a \sqrt{2 \varepsilon}$. Integrating by parts, we arrive at the  Mukhanouv-Sasaki action \cite{MS}, which describes a canonically normalized scalar field with time dependent mass $\frac{Z^{\prime\prime}}{Z}$, where prime denotes the derivative with respect to conformal time $dt=a d\eta$. We can thus find the equation of motion for $\nu$ in momentum space, as:
  \bea \label{nuEquatInMomentun} \nu_k^{\prime\prime}+ ( k^2-\frac{Z^{\prime\prime}}{Z})\nu_k=0\eea
  In the quasi-de-Sitter limit, where $\varepsilon$ changes slowly, the expression $\frac{Z^{\prime\prime}}{Z}$ reduces to $\frac{2}{\eta^2}$, and by restricting the solution to be of positive frequency, which corresponds to the standard Bunch-Davis vacuum \cite{BD} deep inside the Hubble scale, we find the solution of the mode equation (Eq. \ref{nuEquatInMomentun}) as:
  \bea \label{nuSolution} \nu_k = \frac{\exp(-i k  \eta)}{\sqrt{2 k}}(1-\frac{i} {k \eta} )\eea

    Now, the quantization of the cosmological scalar perturbations is performed via a standard procedure in the Heisenberg picture, where the field $\nu$ and its conjugate momentum $\pi$ are promoted to operators $\hat{\nu}$ and $\hat{\pi}$ with equal-time commutation relations. Then each fourier mode in Eq. (\ref{nuSolution}), which represents an independent harmonic oscillator with time dependent frequency, is expressed via creation $\hat{a}_k$ and annihilation $\hat{a}_k^\dagger$ operators:
\bea \label{nuInTermsOfquantumOperatro} \hat{\nu}_k=\nu_k(\eta) \hat{a}_k + \nu^\star_k {\hat{a}_{-k}}^\dagger ,  \hat{\nu}(\eta,x)=\int \frac{d^3k}{(2\pi)^3}[\nu_k(\eta) \hat{a}_k + \nu^\star_k {\hat{a}_{-k}}^{\dagger} ] \exp(ikx)\eea
with the commutation relations
\bea \label{commutiaonRelaiton} [\hat{a}_{k},{\hat{a}^\dagger}_{k^\prime}]=(2\pi)^3\delta^3(k-k^\prime) &,& [\hat{a}_{k},\hat{a}_{k^\prime}]=[{\hat{a}^\dagger}_{k},{\hat{a}^\dagger}_{k^\prime}]=0\eea

The two-point correlation function, some time after the Hubble radius crossing, is given by the VEV $<0|\nu_{k}(\eta) \nu_{k^\prime}(\eta) |0>$ at $\eta=\eta_i \equiv \eta(t_i) = - \frac{1}{H a_i}$. We define the power spectrum as
 \bea <0|\hat{\nu}_k(\eta_i)\hat{\nu}_{k^\prime}(\eta_i)|0>&=& \frac{2\pi^2}{k^3}\Delta^2_\nu(k) (2\pi)^3\delta^{(3)}(k+k^\prime) \eea
 Using the solution of Eq.(\ref{nuSolution}), we obtain, on superhorizon scales $(\frac{1}{k}) >> \frac{1}{aH})$, with $\eta \sim \frac{-1}{Ha}$ in the deSitter background, and so $\frac{ k}{aH}=|k\eta|<<1 $, the expression
 \bea \Delta^2_\nu(k)&=& \frac{1}{4 \pi^2 \eta^2} \approx \frac{H^2 a^2}{4\pi^2 },\eea
 whence the dimensional power spectrum of $\Re$ at the time of horizon crossing $ k =a H$ is
\bea \Delta^2_\Re(k)= \Delta^2_\nu(k)/Z^2 &=&\frac{H^2}{8 \pi^2  \varepsilon}.\eea
Using $(d\ln k = d \ln (aH) \sim Hdt)$ and $(d\ln \Delta^2_\Re = 2 dH/H - d\varepsilon/\varepsilon)$, we deduce thus the ``scalar'' spectral index of $\Re$:
\bea \label{SpectralIndexScalar} n_s -1 = \frac{d\ln \Delta^2_\Re}{d \ln k}|_{k=aH}&=& \frac{2 \dot{H}}{H^2}-\frac{\dot{\varepsilon}}{H\varepsilon}= -2 \varepsilon - \delta \approx -4\ell^2.\eea

 \subparagraph{\bf \underline{Tensor Perturbations}} Similarly to the scalar perturbations, the tensor spectrum is generated by quantum vacuum fluctuations which freeze out at horizon crossing.  Considering a flat FRW background plus tensor perturbations denoted by $h$, we have the metric
 \bea \label{pertTensorFRW} ds^2=[-dt^2+a^2(t)(\delta_{ij}+2h_{ij})dx^i dx^j]\eea
 with $g^{ij} h_{ij}=h^i_{i}=0$. We obtain a quadratic action for tensor modes by expanding the Einstein-Hilbert action to second order in tensor perturbations, and we get:
 \bea \label{TenorExpandedAction} S_T=\int  dt d^3x \frac{a}{4}  [a^2 \dot{h}^{2}_{ij}-(\partial_k h_{ij})^2]\eea
The symmetric, transverse and traceless conditions on $h_{ij}$ leave two physical degrees of freedom which may be parameterized by two fixed polarization tensors $e_+$ and $e_\times$. We follow \cite{linde} and define the fourier expansion as
\bea \label{fouriehModeOfTensorPert} h_{ij}=\int \frac{d^3k}{(2\pi)^3}\Sigma_{s=+,\times}\phi_{ij}^s(k) h_k^s(\eta) \exp {ikx}\eea
with $k^i \phi_{ij}=\phi_{ii}=0$ and $\phi^s_{ij}\phi^{s^\prime}_{ij}=2\delta_{ss^\prime}$. We also define the variable $\mu^s_k=Z_T h ; Z_T=a/\sqrt{2}$ in terms of which the action is of canonical form. Then
\bea \label{TensoractionInTerOfmu} S=\Sigma_{s=+,\times} \half \int d^3k d\eta [(\mu^{s\prime}_k)^2-(k^2-\frac{Z^{\prime \prime}}{Z}((\mu^{s}_k)^2)] \eea
The mode equation is given by
\bea \label{TesorModeEquation} \mu^{s\prime\prime}_k+(k^2-\frac{Z^{\prime\prime}}{Z})\mu^{s}_k=0\eea
The modes $\mu_k$ are quantized in the same way as we quantized the scalar modes, and at horizon crossing, where modes freeze out, we have the dimensional power spectrum
\bea \Delta^2_h(k)&=&\frac{2 H^2}{\pi^2}. \eea
We may also define a spectral index of tensor perturbations $n_T$
\bea \label{spectralIndexOfTesor} n_T \equiv \frac{d ln \Delta^2_h}{d lnk}&\approx& -2 \varepsilon = -4\ell^2\eea

 \subparagraph{\bf \underline{$r$ versus $n_s$}}
For times well before the end of inflation $\varepsilon<<1$, when both $\Delta^2_h(k)$ and $\Delta^2_\Re(k)$  remain approximately constant, we can estimate the tensor-to-scalar ratio as
\bea \label{scalarToTensorRatio} r \equiv \frac{\Delta^2_h(k)}{\Delta^2_\Re(k)} &=& 16 \varepsilon =32\ell^2\eea

Thus, from Eqs. (\ref{SpectralIndexScalar},\ref{scalarToTensorRatio}), we deduce the following straightline relation
\bea
\label{exponential_law_straightline}
r &=& 8 (1-n_s)
\eea

Actually, this relation characterizes an exponential potential law, which is anticipated as by expressing the action of Eq. $(\ref{action})$ in terms of a field $\chi$ with canonical kinetic term, we have \bea
\chi =\frac{ \ln (\phi)}{\ell} \Rightarrow V(\chi) \propto e^{-2 \ell \chi}. \eea
Applying now the well known formulae in the ``potential method":
\bea
\label{pot-meth}
\varepsilon = \frac{1}{2}\left(\frac{V'}{V}\right)^2 &,& \delta = 2\left[\left(\frac{V'}{V}\right)^2-\frac{V''}{V}\right],
\eea
we get exactly the same results as in the `systematic method'.
We plot in Fig. \ref{fig1} this straightline of Eq. (\ref{scalarToTensorRatio}) for the choice $\ell \in [0.1,0.2]$, and we see that the power law inflationary model
can not accommodate recent data even for a sub-Planckian length scale $\ell$, as it lies outside the acceptable observable region.

\section{Variant with non-minimal coupling to gravity (non-MCtG)}
We seek now a variant of the model where we introduce a non-MCtG, so that to start with the action:
\bea S= \int d^4x \sqrt{-g} \lbrace\frac{R}{2}  -\frac{1}{2} f(\phi) g^{\mu \nu} \partial_\mu \phi \partial_\nu \phi -V(\phi) - \frac{1}{2} \xi R \phi^2 \rbrace \eea
where $\xi$ is a coupling constant expressing the coupling between gravity represented by $R$ and the field $\phi$ representing the variation of the strong coupling constant.

As explained in the introduction, we shall adopt two formalisms which, unlike the MCtG case, become distinct physically upon introducing the $\phi^2 R$ term: the ``Metric" formulation where the connection is assumed to be Levi-Civita and is not independent of the metric, and the ``Palatini" formulation where we treat the Christoffel symbols as independent of the metric. This difference is reflected at the first stage of finding the background equations in the ``systematic approach" upon varying the metric: the Riemann tensor remains fixed in ``Palatini" unlike ``Metric", so we get:
\bea \label{EinsteinEqs}
\mbox{``Metric":}\\
G_{\mu\nu} (1 -\xi \phi^2) &=&  \gmu \left\{(-\half f+2\xi) (\partial \phi)^2 -V \right\}  +(f-2 \xi)\partial_\mu \phi \partial_\nu \phi - 2\xi \phi (\nabla_\mu \nabla_\nu - \gmu \square)\phi \nonumber \\
\mbox{``Palatini":} \label{EinsteinEqs2}\\
G_{\mu\nu} (1 -\xi \phi^2) &=&  \gmu \left\{-\half f (\partial \phi)^2 -V \right\}  +f\partial_\mu \phi \partial_\nu \phi \nonumber
  \eea
where again $G_{\mu\nu}$ is the Einstein tensor, $\square \phi \equiv g^{\rho \sigma} \nabla_\rho \nabla_\sigma \phi$ with  $\nabla_\mu \nabla_\nu \phi=\partial_\mu \partial_\nu \phi - \Gamma^{\lambda}_{\mu\nu}\partial_{\lambda}\phi$ where $\Gamma^{\lambda}_{\mu\nu}$ denote the Christoffel symbols. Assuming a homogenous and isotropic universe with $\phi=\phi(t)$, we have $\square \phi = -\ddot{\phi}-3H\dot{\phi}$.

Again, one can find the modified Friedmann equations by taking the $00$ and $11$ component of the equations (\ref{EinsteinEqs}, \ref{EinsteinEqs2}) to get:
\bea \label{1st-friedmann-eqs} \mbox{``Metric":}\\
3(1 - \xi \phi^2)H^2=\half f \dot{\phi}^2+V+6\xi H \phi \dot{\phi} &,&
(1 - \xi \phi^2) \dot{H} + \frac{1}{4} f \dot{\phi}^2=\xi \left( \dot{\phi}^2 - H \phi \dot{\phi} + \phi \ddot{\phi} \right)\nonumber \\
\mbox{``Palatini":}\\
3(1 - \xi \phi^2)H^2=\half f \dot{\phi}^2+V &,&
(1 - \xi \phi^2) \dot{H} + \frac{1}{4} f \dot{\phi}^2=0 \nonumber
 \eea
 In addition, within ``Palatini formulation", one should determine the connection $\Gamma$ dynamically. However, since the pure gravity term is just an $R$-term then $\Gamma$ reduces to Levi-Civita connection \cite{Demir-0803.2664}, and the difference between ``Metric" and ``Palatini" methods comes from the matter sector reflected in the different forms of the bacjground \& perturbed equations in the ``systematic method".

As to the equation of motion for $\phi$, it is the same in both ``Metric" and ``Palatini" approaches:
\bea \label{eqofmotionforphi} f(\phi) \ddot{\phi}+3 H f(\phi) \dot{\phi} +\frac{1}{2} f_{,\phi} \dot{\phi}^2 +  V_{,\phi}+ \xi R \phi =0.\eea
One can now follow exactly the procedure detailed in section \ref{1} in order to compute the spectral parameters in the ``systematic method''.

As to the ``potential method", one can not use directly the standard formulae for the spectral parameters involving $V$ and its derivatives here, as we have the extra term $\xi R \phi^2$. However, there is an alternative way where one transforms the action, by rescaling the metric, from the original physical ``Jordan'' frame to a new non-physical ``Einstein'' frame similar in form to the action without extra terms. One can then apply the ``shortcut'' standard formulae for the transformed potential.

The difference between the ``Metric" and ``Palatini" formalisms in evaluating the physical parameters in the ``potential method'' does not originate from a different `potential'. Actually, the potential in `Einstein' frame, in terms of the `Jordan' field $\phi$, is the same in both formulations, however the new `canonical' field $\chi$ is expressed in terms of $\phi$ differently in the two formulations. Let us clarify this as follows. We start from the action in Jordan frame:
\bea
\label{NMCtG-action}
S_J &=& \int d^4x \sqrt{-g} \left[ \frac{1-\xi \phi^2}{2} R - \frac{1}{2} f (\partial \phi)^2 -V\right]
\eea
Transforming the metric conformally
\bea \label{conformal}
g_{\m\n} \rightarrow \tilde{g}_{\m\n} = \Omega^2 g_{\m\n} &:& \Omega^2=1-\xi \phi^2,\
\eea
we get the action in Einstein frame. Again, changing the metric as above leads to a difference between ``Metric" and ``Palatini" formulations because the $R$-term changes differently in the two formalisms. Although we get the same ``form" of the action \cite{deFelice}:
\bea
\label{NMCtG-action-E}
S_E &=& \int d^4x \sqrt{-\tilde{g}} \left[ \frac{\tilde{R}}{2}  - \frac{1}{2}  \tilde{g}^{\m\n}\partial_\m \chi \partial_\n -U(\chi)\right]
\eea
in the two methods with:
\bea
\label{UNMCtG}
U(\chi) &=& \frac{V(\phi)}{\Omega^2} = \frac{V(\phi)}{1-\xi \phi^2}
\eea
 however the `canonical' field $\chi$ is given differently in the two methods by :
\bea
Z^{-1}\equiv \left(\frac{\partial \chi}{\partial \phi}\right)^2 &=& \frac{\Omega^2+6\xi \ell^2 \phi^4}{\ell^2 \phi^2 \Omega^4} \mbox{ ``Metric"} \label{Zmetric}\\
&=& \frac{1}{\ell^2 \phi^2 \Omega^2} \mbox{        ``Palatini"} \label{ZPalatini}
\eea
In Einstein frame, the field $\chi$ is minimally coupled to $\tilde{R}$, and has a canonical kinetic energy term, so one can apply directly the folrmulas (cf. Eqs. \ref{pot-meth}):
\bea
n_s = 1 - 6 \varepsilon + 2 \eta &,& r= 16 \varepsilon, \\
\label{eps-eta-Potential}
\varepsilon = \frac{1}{2} \left[\frac{\frac{dU}{d\chi}}{U}\right]^2
&,&
\eta = \frac{\frac{d^2U}{d\chi^2}}{U}=2 \varepsilon -\frac{\delta}{2}
\eea
Note here that, unlike the ``Metric" case (Eq. \ref{Zmetric}), one can integrate and invert (Eq. \ref{ZPalatini}) to get
\bea
\label{closedPalatini} \chi^{\mbox{\tiny Pal}} = -\frac{\tanh^{-1} \sqrt{1-\xi \phi^2}}{\ell} &\Rightarrow& U^{\mbox{\tiny Pal}}(\chi) = V_0 \xi \coth^4(\ell \chi)
\eea
Thus, in the ``Palatini" formalism, one can plot the `canonical' potential, and investigate directly whether or not it has a `flat' portion on which slow-roll inflation takes place.

In general, we can express $\varepsilon, \eta$ in terms of the old field, for example
\bea
\label{epsOld}
\varepsilon &=& \frac{1}{2} \left[\frac{dV}{d\phi}+V\frac{4\xi \phi}{1-\xi \phi^2}\right]^2 \frac{Z}{V^2}
\eea and similarly for $\eta$.

Both the ``systematic method'' and the ``potential method'' give the following results in terms of $y=\xi \phi^2$:
 \bea \label{nsAndr}\mbox{``Metric":   }\\ n_s&=& \frac{(1-y)^2 + 12 \ell^4 y^2 (2+ 8 y - 15 y^2) + 4 \ell^2 (-1 +12 y -20 y^2 + 9 y^3)}{(1-y + 6 \ell^2 y^2)^2} \nonumber \\
 r&=&\frac{32 \ell^2 (1-3y)^2}{1-y + 6 \ell^2 y^2} \nonumber \\
\mbox{``Palatini":    } \\
 n_s &=& \frac{1-y-4 \ell^2 (1-11y+12y^2)}{1-y}
\nonumber \\
r&=&\frac{32 \ell^2 (1-3y)^2}{1-y } \nonumber
   \eea
Remember that $\phi$ here denotes the field values at the start of inflation, and so one can scan over these `initial' values with the other two unknown parameters $\ell, \xi$ in the model in order to test the model. However, one can  study directly the relation between $n_s$ and $r$ and show that the model is not viable.

Actually, expanding $\ell$ around $L$, where $L$ is a given nominal value close to $\ell$ so one can neglect terms of order $O(\ell - L)$,  we get:
\bea \label{nsAndrExpandL} \mbox{``Metric":}\\
 1-n_s&=& \frac{4L^2(1-12 y + 23 y^2 -6 L^2 y^2 -12 y^3 -24 L^2 y^3 +54 L^2 y^4)}{(1-y+6y^2 L^2)^2}+O(\ell-L)\nonumber \\
 r&=& \frac{32L^2(1-3y)^2}{1-y+6L^2y^2} +O(\ell-L) \nonumber \\
\mbox{``Palatini":}\\
 1-n_s &=& \frac{4 L^2 (1-11y+12y^2)}{1-y} +O(\ell-L) \nonumber
\\ r &=& \frac{32L^2(1-3y)^2}{1-y} +O(\ell-L)\nonumber
\eea
We see directly that for $L$ not too small, we need $y \sim 1/3$ in order to make $r$ small near its experimental value, but then we get $1-n_s \sim (-8L^2)(\times \frac{1}{(1+ L^2)})<0 $ in ``Palatini" (``Metric") formulation, which is rejected experimentally. Thus we need short length scale $\ell <<1$.

Although it is difficult to describe within quantum field theory length scales shorter than Planck length ($L_{pl}$), however the question whether or not $L_{pl}$ is the shortest length scale appearing in any physical theory or rather the smallest measurable such length, due to creation of black holes when carrying out such a measurement, is still not clear, and one can argue within string theory or loop quantum gravity that basic entities are of length comparable to Planck Length\cite{Bachas,Antoniadis}. Taking $\ell$ too small puts conditions on the `initial' speed of the physical field $\dot{\phi}$ (look at Eq. \ref{slowrollparametr}) which may be required to be fine tuned in order to maintain  $\varepsilon = \frac{\dot{\phi}^2}{\ell^2 \phi^2 H^2}$ small, and idem for $\eta$. The same argument applies in Eq. (\ref{equMotionOf}) where the combination $f\dot{\phi}^2=\frac{\dot{\phi}^2}{\ell^2 \phi^2}$ should remain small.

\begin{figure}[H]
\includegraphics[width=15.5 cm]{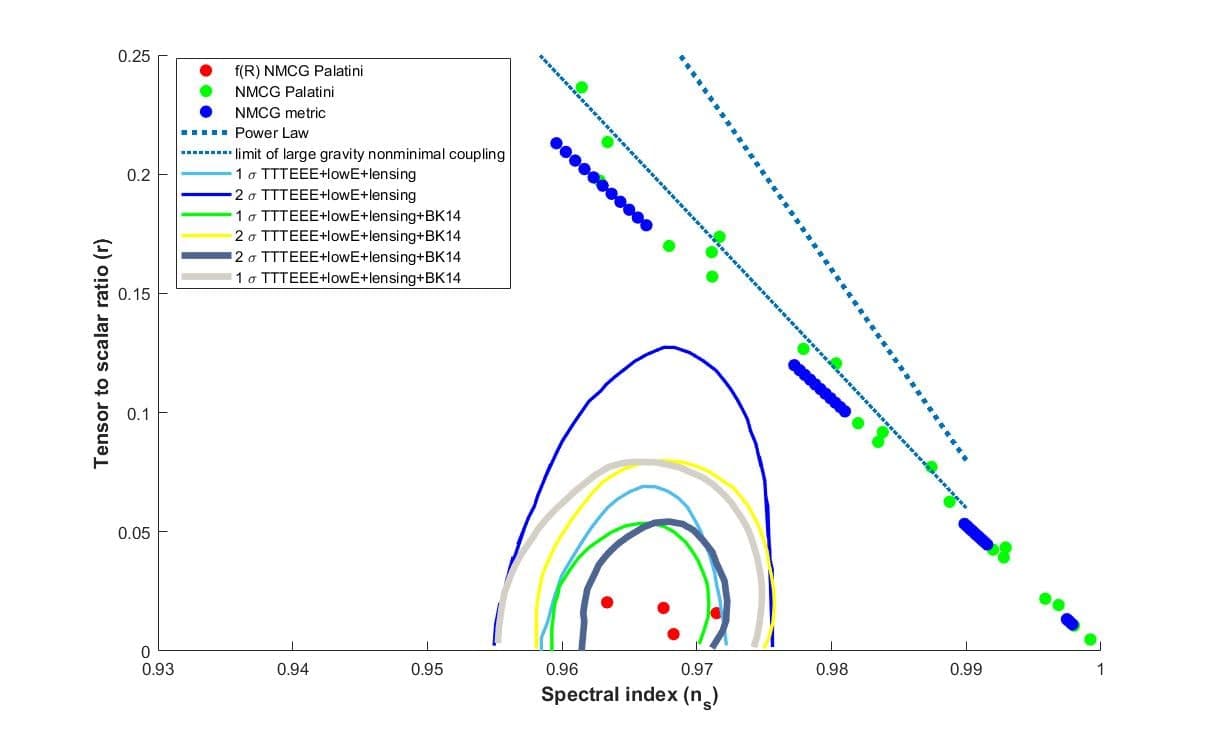}
\caption{ Variant model of nonminimal coupling to gravity contrasted to Planck 2018, separately or combined with other experiments, contour levels of spectral parameters ($n_s,r$). Both limits of vanishing and very strong nonminimal gravity coupling $\xi$ are not viable, with slope $r/(1-n_s)=8$ and $6$ respectively. Green (Blue) dots correspond to ``Palatini" (``Metric") non-MCtG variants, whereas acceptable red points correspond to $f(R)$ with non-MCtG variant. \label{fig1}}
\end{figure}

However, even if we discard temporarily the Bekenstein condition that the model should not contain a length scale shorter than Planck length, we still find that the model for small $\ell$ is not viable either. Actually, for $\ell <<1$ we get
\bea \label{non-MCtG-slope}
\frac{r}{1-n_s} &=& \frac{8(1-3y)^2}{1-11y+12y^2} \mbox{``Palatini"} \left(+O(\ell^2)\mbox{``Metric"} \right)
\eea
In the limit $\ell<<1, |y|>>1$ (which means physically a strong non-MCtG), we get
\bea \label{nsAndrExpandCase1}
r &=&  6 (1-n_s)+O(\ell^2,\frac{1}{y})
\eea
We see directly that the experimentally rejected straightline, corresponding to the exponential potential, with slope $-8$ (cf. Eq. \ref{exponential_law_straightline}) is replaced by a straightline with slope ($-6$), approaching thus the experimentally allowed region but still not intersecting it, as Fig. \ref{fig1} shows. Outside the strong non-MCtG limit, one cannot either reach the data, since the $r/(1-n_s)$ in (Eq. \ref{non-MCtG-slope}) attains its minimal value of around $5.25$ while the data impose $r/(a-n_s) < 4$.  Fig. \ref{fig1} shows the results of scanning the model parameters within their corresponding ranges
\bea
\mbox{``Palatini":} &&
\ell \in [0.01, 0.1],
\phi\in [0.1,1.0],
\xi \in [-10.5,-9.5];\\
\mbox{``Metric":} &&
\ell \in [0.01,0.1],
\phi= [0.9,1.0],
\xi=[-0.6,-0.5];
\eea
where we see that the ``Palatini" green dots and the ``Metric" blue dots are situated near the straightline $r=6(1-n_s)$ in an excluded region.

The morale from this discussion is that introducing non-MCtG improves the model regarding accommodation of experimental data, but still cannot make it viable. We need an additional degree of freedom, which would be $f(R)$ gravity.

\section{Variant with $f(R)$ gravity with non-MCtG}
We study in this section whether or not the ingredient of $f(R)$ gravity can make the model viable. We start by absorbing the function $f$ into the field to get a `canonical' kinetic energy term. Thus we start with the action:
\bea
\label{fr-action}
S= \int d^4x \sqrt{-g} \left[\frac{1}{2} F(R) + \frac{1}{2} G(h) R - \frac{1}{2} g^{\a\b} \partial_\a h\partial_\b h - V(h)\right]
\eea
where
\bea
\phi=\exp(\ell h),V(h)=V_0 \exp(-2 \ell h), G(h)= -\xi \exp(2 \ell h), F(R) = R + \a R^2
\eea
Again, one can use the ``systematic method" here in both ``Metric" and ``Palatini" formulations. The new $\a R^2$ would contribute involved terms upon changing the metric ($g_{\m\n} \rightarrow g_{\m\n} + \d g_{\m\n}$). We follow \cite{faraoni-0805.1726, Enckel} and introduce an auxiliary field $\psi$ and an action:
 \bea
\label{fr-action2}
S= \int d^4x \sqrt{-g} \left[ \frac{1}{2} G(h) R + \frac{1}{2} \left\{F(\psi) + F'(\psi) (R-\psi)\right\}  - \frac{1}{2} g^{\a\b} \partial_\a h\partial_\b h - V(h)\right]
\eea
The equation of motion of $\psi$ gives $R=\psi$ as long as $F''(\psi) \neq 0$. We change variable $\psi \rightarrow \varphi$ such that ($\varphi=F'(\psi)= 1+2\a \psi$), so we get
\bea
\label{fr-action3}
S= \int d^4x \sqrt{-g} \left[ \frac{1}{2} \left\{\varphi + G(h)\right\} R - \frac{1}{2} \left\{\psi \varphi -F(\psi(\varphi)) \right\}  - \frac{1}{2} g^{\a\b} \partial_\a h\partial_\b h - V(h)\right]
\eea
Transforming the metric conformally
\bea
\label{fr-conformal}
g_{\a\b} \rightarrow \Upsilon^2 g_{\a\b}= \tilde{g}_{\a\b} &:& \Upsilon^2= \varphi+G(h)
\eea
then we get in the ``Metric" formulation \cite{Budhi}:
\bea
\label{fr-action-metric}
S^{\mbox{\tiny ``Metric"}} &=& \int d^4x \sqrt{-\tilde{g}} \left[ \frac{1}{2} \tilde{R} - \frac{3}{4} \frac{\tilde{g}^{\m\n}}{(\varphi+G(h))^2} \nabla_\m (\varphi + G(h)) \nabla_\n (\varphi + G(h)) \nonumber \right. \\ && \left.    -  \frac{1}{2} \frac{1}{\varphi+G(h)} \tilde{g}^{\a\b} \partial_\a h\partial_\b h - \tilde{V}(h,\varphi)\right] \\
\tilde{V}(h,\varphi)&=& \frac{V(h)+W(\varphi)}{(\varphi + G(h))^2}
\label{tildeV}
\eea
where
\bea \label{W} W(\varphi)&=& \frac{1}{2} \left[ \psi \varphi - F(\psi(\varphi))\right]
\eea
 We see that in the ``Metric" formulation, we get a kinetic energy term for ($\varphi + G(h)$), and the field $\varphi$ is dynamic, i.e. its equation of motion cannot be solved algebraically.

 For simplicity, then,  we restrict the study from now on to the ``Palatini" formulation, remembering that $R$ is not Levi-Civita connection since the pure gravity is not represented by a simple $R$-term. Under this formulation, we get:
 \bea
\label{fr-action-palatini}
S^{\mbox{\tiny ``Palatini"}} &=& \int d^4x \sqrt{-\tilde{g}} \left[ \frac{1}{2} \tilde{R}    -  \frac{1}{2} \frac{1}{\varphi+G(h)} \tilde{g}^{\a\b} \partial_\a h\partial_\b h - \tilde{V}(h,\varphi)\right]
\eea
where again $\tilde{V}(h,\varphi)$  is given by Eq. (\ref{tildeV}), and where eq. (\ref{W}) is again valid.

We see here that $\varphi$ is not a new degree of freedom, since its equation of motion can be solved algebraically and gives:
\bea
\label{varphi}
\varphi &=& \frac{1+G(h)+8\a V(h) +2 \a G(h) (\partial h)^2}{1+ G(h) -2\a (\partial h)^2}
\eea

Substituting (Eq. \ref{varphi}) in (Eq. \ref{fr-action-palatini}), we get (dropping the ``Palatini'' superscript and the $\sim$ over the metric):
\bea
\label{fr-action-palatini-almostfinal}
S &=& \int d^4x \sqrt{-g} \left[ \frac{R}{2} -  \frac{1}{2} \frac{1}{(1+G(h))(1+8\a \bar{U})} g^{\a\b} \partial_\a h \partial_\b h \right. \nonumber \\ && \left.
+\frac{\a}{2} \frac{1}{(1+G(h))^2(1+8 \a \bar{U})} (\partial^\a h \partial_\a h)^2
- \frac{\bar{U}}{1+8 \a \bar{U}}\right]
\eea
where
\bea
\label{barU}
\bar{U} &=& \frac{V(h)}{(1+G(h))^2} = \frac{V_0 \exp{(-2 \ell h)}}{(1-\xi \exp{(2 \ell h)})^2}.
\eea
Upon redefining the field ($h \rightarrow \chi$) in order to get a `canonical' kinetic energy term by
\bea
\label{hoverchi}
\frac{dh}{d\chi} &=& \pm \sqrt{(1+G(h)) (1+8\a \bar{U})}
\eea
 we get finally
\bea
\label{fr-action-palatini-final}
S &=& \int d^4x \sqrt{-g} \left[ \frac{R}{2} -  \frac{1}{2} g^{\a\b} \partial_\a \chi \partial_\b \chi
+\frac{\a}{2} (1+8 \a \bar{U}) (\partial^\a \chi \partial_\a \chi)^2
- U\right]
\eea
where
\bea
\label{Uchi}
U &=& \frac{\bar{U}}{(1+8\a \bar{U})}
\eea
We see here that the effect of $\a R^2$-term is reflected in two effects. First, it helps in getting a `flat' effective potential $U$. In fact, irrespective of the form of $\bar{U}$, which can increase indefinitely, we see that the effect of $\a R^2$ is to divide the potential $\bar{U}$ so to get an effective potential showing a flat portion ($U \sim (8\a)^{-1}$) where $\bar{U}$ increases indefinitely. Second, the $\a R^2$-term leads to the appearance of squared kinetic energy $(\partial^\a \chi \partial_\a \chi)^2$. At this stage, one can apply the ``systematic method" in order to treat this term which, upon perturbing the metric $(g^{\m\n} \rightarrow g^{\m\n} + \d g^{\m\n})$, would contribute with terms of first order in metric perturbation that the ``potential method" fails to investigate.

However, for $\a$ small one can neglect this term, since the $(\a \d g)$ would give higher order terms, whereas the $\a (\partial^\a \chi \partial_\a \chi)^2$ would give, in the slow-roll inflationary era, contributions of order $\a \dot{\chi}^4$ which is subdominant compared to the $\a$-correction in $U$. Thus, in this situation, one can apply the ``potential method'' using $U$ as an effective potential and we get (again denoting by $y$ the quantity $\xi \phi^2 = \xi \exp(2 \ell h)$, and by $\b$ the combination $\xi \a V_0$)
\bea
\label{fr-ns-r}
1-n_s = \frac{4 \ell^2 (1-11y+12y^2)}{1-y} &,& r=\frac{32 \ell^2 y (1-3y)^2 (1-y)}{y+8 \b -2y^2 +y^3}
\eea
We see that $n_s$ is exactly the same as in non-MCtG Palatini scenario, whereas $r$ is different, and  even though $\a$ should be small for the ``potential method'' to work here, however we have freedom in $\b= \a \xi V_0$.

One can take $\ell$ large as long as one adjusts $y$ to be around $(0.1)$ or $(0.8)$ (the roots of $1-11y+12y^2=0$ in the numerator of $1-n_s$). Adjusting $\b$ to be large so that to make $r$ and $\frac{r}{1-n_s}$ small at will, we can get $0<1-n_s<<1, 0<r<<1, r/(1-n_s)<<4$, and the model is viable.

Fig. \ref{fig1} shows that there are acceptable points, colored in red, for the following scanning:
\bea
\label{scanning}
\ell \in [1,1.5], y=\xi \phi^2 \in [0.10,0.115], \b = \a \xi V_0 \in [5,25].
\eea

\section{Variant with two-Exponentials potential}

 We follow here \cite{chamounJCAP}, and consider two gauge groups $G_k$ ($k=1,2$) with varying coupling constants $g_k(x)=g_{0k}\phi_k(x)$, where $g_{0k}$ represents the usual energy-dependent part whereas the non-canonical fields $\phi_{k}(t)$ encode the temporal variation. Again, Gauge invariance gives the corresponding Lagrangian kinetic energy parts $\frac{1}{\ell^2_k \phi_k^2} g^{\mu \nu} \partial_\mu \phi_k \partial_\nu \phi_k$, whereas the potential is derived from the corresponding condensates ($\frac{\langle G^2_k\rangle}{\phi_k^2}$).  Redefining the canonical fields $\chi_k \propto \frac{\log \phi_k}{\ell_k}$, we get the potential:
    \bea \label{potentialTwoFields} V(\chi_1, \chi_2) &=& \langle G^2_1 \rangle \left(e^{-2\ell_1\chi_1} + \mu e^{-2\ell_2\chi_2}\right)\eea with $\mu$ the ratio between the two condensates.
 We have here a two-field inflation ($\chi_1,\chi_2$), and as we have no information about the relative strength of the different condensates, let alone a non-perturbative theory about them, we may treat $\mu$ as a free parameter. With extra degrees of freedom corresponding to ($\ell_1, \ell_2, \mu$) and the initial values of the fields ($\chi_1, \chi_2$), accommodating the cosmological data should be possible.

\begin{figure}[H]
  \center
\includegraphics[width=7.5 cm]{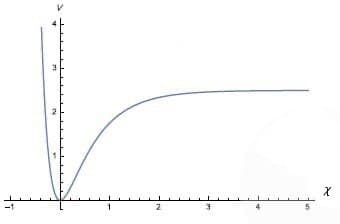}
\caption{ \label{curve} The potential of eq. (\ref{twopotV}), with ($\ell_1 > \ell_2$).}
\end{figure}

\begin{figure}[H]
\includegraphics[width=15.5 cm]{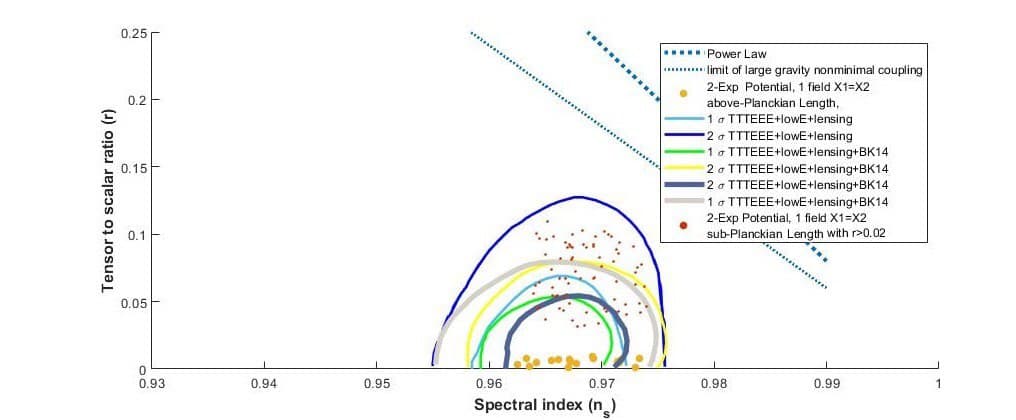}
\caption{ Contour levels, for $r$ versus $n_s$, of Planck 2018  experiments and their Combined analyses with other experiments. \label{fig2} Predictions of single inflaton field are shown. The upper (lower) straightline represents an exponential potential (with very large gravity nonminimal coupling and subPlanckian length scale) which clearly does not accommodate data. However, the brown and chrome dots represent a variant to the potential involving two exponential terms, which can accommodate the data. }
\end{figure}

 Alternatively, one can look at the two-fields specific trajectory as an effectively single inflaton field ($\chi$).
 We adopt this procedure and seek a variant mimicking as much as possible the standard inflation paradigm with a plateau corresponding to slow rolling and a minimum at which the inflation ends. For this, let us assume, as a toy model with no attempt of justification, that the variation of the coupling constants of the two groups is such that we have ($^{\ell_k}\sqrt{\frac{g_k}{g_{0k}}}$) is identical for both groups, so we have $\chi_1=\chi_2$, and so by taking a canonical field $\chi = \frac{\chi_1}{\sqrt{2}}=\frac{\chi_2}{\sqrt{2}}$ we get a potential with two exponentials:
\bea \label{2potV}
V(\chi) &\propto& \left(e^{-\sqrt{2}\ell_1\chi} + \mu e^{-\sqrt{2}\ell_2\chi}\right) \eea
A priori, this is a $3$-parameter potential. However, imposing that the inflation ends when the temporal variation fields ($\phi_k = e^{\ell _k \chi_k}$) reach their today's value equal, by convention, to $1$, we seek a variant such that the minimum occurs at ($\chi = 0$). Then $\mu$ is no longer a free parameter but equal to ($- \frac{\ell_1}{\ell_2}$), or, alternatively, it is `fine-tuned' in order to get a desired form suitable for slow roll inflation. Shifting also the potential upward so that it corresponds to zero cosmological constant we have finally the potential:
\bea \label{twopotV}
V(\chi) &=& M^4 \left[e^{-\sqrt{2}\ell_1\chi} -\frac{\ell_1}{\ell_2} e^{-\sqrt{2}\ell_2\chi} + \left(\frac{\ell_1}{\ell_2} - 1\right)\right] \eea
 where $M$ is a mass scale proportional to the $G_1$ condensate.
 For $\ell_1 > \ell_2$ we have the desired form depicted in (Fig. \ref{curve}), and no need to fine tune $\ell_1, \ell_2$ in order to fit the data. We stress again that we present here just a toy model aiming to show a possible link between variation of constants and the inflationary scenario\footnote{Actually, for $\mu<0$, the two-inflaton fields potential of Eq. (\ref{potentialTwoFields}) is not bounded from below. Although this makes any minimum prone to decay via tunneling and breaks stability, however one expects new physics to settle in ending the inflation before this tachyonic instability takes place.}.
We see in Fig. \ref{fig2} that this variant still fits the Planck 2018 data, separately and combined with other experiments. The brown thin (chrome thick) dots in Fig. (\ref{fig2}) corresponding to $\ell_1, \ell_2<L_{pl}$ ($\ell_1, \ell_2>L_{pl}$) show that such a potential can accommodate the data.

\section{Summary and Conclusions}
In this letter, we rederived  the exponential inflationary model generated by a varying strong coupling constant following the ``systematic'' procedure leading to the expressions of the spectral index for scalar $n_s$ and tensor $n_T$ perturbations, and the scalar-to-tensor ratio $r$.
We single out the origin of the difference between ``Metric" \& ``Palatini" formulations in both the ``systematic method" and the shortcut ``potential method". When applied to
 a variant of the model assuming non-MCtG, we found that such a variant can not survive comparison with observational data. To remedy this, we add the ingredient of $f(R)$ gravity to the model, and for simplicity we adopted the ``Palatini" formulation. We found that the effect of $\a R^2$-term would show itself in `Einstein' frame by two terms: the appearance of kinetic energy squared that can be dropped when $\a$ is small allowing the use of ``potential method", provided we take an effective potential, representing the second effect of $\a R^2$, which can show a desired form. Fitting with data, albeit it requires adjusting the combination $(\xi \phi^2)$, does not, however, need fine tuning for $(\ell, V_0, \a)$, and is natural even with $\ell > L_{pl}$. This comes because $n_s$ remains like the $R$-gravity case, whereas $r$ can be divided at will.   We also showed that another variant of the model with two exponential terms in the potential can accommodate Planck 2018 data with other experiments for both cases ($\ell < L_{pl}$) and $\ell > L_{pl}$.

\vspace{6pt}
\section*{{\large \bf Acknowledgements}}
 The authors thank the anonymous referees for their comments and suggestions. N. Chamoun acknowledges support from ICTP-Associate program and from the Alexander von Humboldt Foundation.



\end{document}